%Paper: astro-ph/9502091
%From: salucci@tsmi19.sissa.it
%Date: Wed, 22 Feb 1995 17:54:09 +0200

\magnification 1200
\def\mincir{\raise -2.truept\hbox{\rlap{\hbox{$\sim$}}\raise5.truept
\hbox{$<$}\ }}
\def\magcir{\raise -2.truept\hbox{\rlap{\hbox{$\sim$}}\raise5.truept
\hbox{$>$}\ }}
\def\minmag{\raise-2.truept\hbox{\rlap{\hbox{$<$}}\raise 6.truept\hbox
{$>$}\ }}
\def\gr{\kern 2pt\hbox{}^\circ{\kern -2pt K}} %  ====> GRADI KELVIN
\parindent 2.truecm
\baselineskip 19pt
\vglue 2.truecm
\centerline {\bf ROTATION CURVES OF 967 SPIRAL GALAXIES }
\bigskip
\bigskip
\centerline {\bf Massimo Persic$^+$ and Paolo Salucci}
\bigskip
\centerline {SISSA, Strada Costiera 11, I-34014 Trieste, Italy}
\centerline {$^+$ and Osservatorio Astronomico, Trieste, Italy}
\bigskip
\bigskip
\leftline {{\it email:} INTERNET: PERSIC@tsmi19.sissa.it;
SALUCCI@tsmi19.sissa.it}
\leftline {$~~~~~~~~$ DECNET: 38028::PERSIC; 38028::SALUCCI}
\vglue 1.truecm
\centerline{{\it Ap.J. Supplement}, in press}

\vfill\eject

\vglue 2.truecm
\parindent 1.truecm

\noindent
{\bf Abstract.} We present the rotation curves of 967 southern spiral galaxies,
obtained by deprojecting and folding the raw H$\alpha$ data originally
published by Mathewson, Ford \& Buchhorn (1992). For 900 objects, we also
present, in figures and tables, the rotation curves smoothed on scales
corresponding to 5\% - 20\% of the optical size: of these, 80 meet objective
excellence criteria and are suitable for individual detailed mass modelling,
while 820, individually less compelling mainly because of the moderate
statistics and/or limited extension, are suitable for statistical studies. The
remaining 67 curves suffer from severe asymmetries, small statistics, and large
internal scatter that may largely limit their use in galaxy structure studies.

The deprojected folded curves, the smoothed curves, and various related
quantities are available via anonymous ftp at galileo.sissa.it in the directory
/users/ftp/pub/psrot.

\vglue 2.truecm

\noindent
{\it Subject headings:} galaxies: kinematics and dynamics --- galaxies:
internal
motions --- galaxies: structure

\vglue 1.5truecm
%\centerline{Ref.: SISSA 115/92/A}
%\vfill\eject

\beginsection 1. Introduction.

Rotation curves (hereafter RCs) are the prime tracers of the mass distribution
of spiral galaxies (see the review by Ashman 1992). As such they represent a
main observable of the whole process of galaxy formation and are relevant for
considerations of the nature of dark matter. A detailed knowledge of their
morphology and their implications for structure is clearly essential for
theories and numerical experiments of galaxy formation (e.g.: Cen \& Ostriker
1993; Navarro \& White 1994; Evrard et al. 1994) as well as for testing
unconventional dynamics/gravitation (e.g., Mannheim 1993). The accuracy of such
knowledge is, of course, crucially increased when large samples of good-quality
curves become available.

The H$\alpha$ RCs of nearly one thousand southern spirals published by
Mathewson, Ford \& Buchhorn (1992; hereafter MFB) represent by far the biggest
database on galaxy rotation available to date (967 objects: 965 shown in their
Fig.3, plus 2 additional ones [416-G20 and N1241] not shown there but similarly
available on file and also contained in their Table 1). However, in the
MFB paper (and
related public files) the H$\alpha$ data are presented only as
measured velocities, which is adequate for MFB's purpose of
deriving rotation amplitudes in order
to determine galaxy distances and peculiar velocities via the Tully-Fisher
relation. In the present paper we make available the actual RCs after folding,
deprojecting and smoothing the raw MFB data. Because of its size, homogeneity,
quality, and spanned range of luminosities and asymptotic velocities, the
sample of RCs which we are presenting will serve as a main database for studies
of
galaxy structure.

We have divided the derived RCs into three subsets (see below for details; cf.
cols. [6] and [12] of Table 1). Set A includes 80 RCs that are smooth,
symmetric, have negligible rms internal error, extend out to at least the
optical radius, and have high and homogeneus radial data coverage. Because of
their objective characteristics, these curves can compare with the best optical
one--slit RCs available in the literature (e.g., Rubin et al. 1985) and are
ideally suited for detailed mass structure modelling. Set B includes 820 RCs of
good quality which, however, have some limitations, such as undersampling,
limited extension (i.e., $R({\rm farthest})<$ 3 disk lengthscales), moderate
asymmetries, non-circular motions, or non-negligible rms internal error. Thus
the RCs in this sample are adequate for the investigation of the DM issue, as
soon as specific methods are employed. Set C includes 67 curves with severe
global asymmetries, scanty sampling, very large rms internal scatter, and
conspicuous non-circular motions. In our opinion these RCs are unsuitable for
studying the mass distribution in spirals and may even be considered for
rejection in the luminosity-velocity diagram.

The plan of this paper is as follows. In section 2 we give details of the
folding and smoothing procedure used for obtaining the RCs from the raw MFB
data. In section 3 we arrange the 967 RCs into the three subsets according to
their quality, we fold the raw curves, and, finally, for the best-quality set
we obtain smooth RCs by binning and averaging the folded data. Section 4
contains a summary and the conclusions. The files containing the deprojected
folded curves, the smoothed curves, and the global properties of each galaxy
relevant to the RCs (heliocentric systemic velocity, separation between
kinematic and photometric centers, inclination angle, optical radius) are
available via anonymous ftp at galileo.sissa.it in the directory
/users/ftp/pub/psrot.

\beginsection 2. Data Analysis.

For details regarding the observations and data acquisition and reduction, the
reader is referred to MFB. Here we describe the basic steps which we have taken
in order to obtain the final RCs. For each RC these include: 1) a selection of
the individual velocity data; and 2) the identification of the kinematical
center about which to fold such data. In addition, 3) we have compared the
radius corresponding to the farthest measured velocity with the optical size in
order to evaluate the extension of each RC relative to that of the visible
matter.

The quality of each velocity measurement is indicated by the cross-correlation
coefficient $\rho$ (given in the public data files) between the observed
H$\alpha$ line profile and the template line profile (see MFB for details). We
have found that a good correlation ($\rho > 0.35$) is essential in order to
ensure a high reliability of the measure. In fact, using differences in
velocity between adjacent points (separation $\leq 0.1 \,R_{opt}$), as a
measure of the intrinsic scatter $\sigma$ of the RCs, we have found that for
good cross-correlations (i.e., $\rho>0.35$) the mean rms error velocity is
small, $\sigma \simeq 9$ km s$^{-1}$, while for poor cross-correlations (i.e.,
$\rho \leq 0.35$) the error is quite significant, $\sigma \simeq 20$ km
s$^{-1}$.

The photometric center representing the maximum emission in the $I$-band does
not necessarily coincide with the dynamical center of the galaxy. Moreover,
MFB's procedure for estimating the systemic (heliocentric) velocity, though
adequate for their purpose, may be too simplified for the raw data to be folded
successfully. In order to obtain the actual coordinate of the kinematical
center we have proceeded as follows: we have assumed that, once folded, each
curve must be perfectly symmetric around its kinematical center of coordinates
$(V_{sys}^{hel}, R_0)$, so that the curve defined by the approaching arm,
$\big| V_a(|R-R_0|)-V_{sys}^{hel} \big|$, coincides within its internal scatter
with the one traced by the receding arm, $\big| V_r(|R-R_0|)-V_{sys} ^{hel}
\big|$. In practice, we have started from MFB's coordinates $(V^{hel\,(
MFB)}_{sys},0)$, which in many cases are excellent solutions; then, if
warranted, we have estimated the actual center by folding the data around
slightly different zero-points until global symmetry of the velocity arms is
reached (i.e., maximized). In most cases this process has quickly converged to
solutions not very far away from the starting ones ($\big| R_0 \big| \leq 2"$;
$\big|V_{sys}^ {hel}-V_{sys}^{hel\,(MFB)}\big| \leq 20$ km s$^{-1}$). The
resulting RCs are generally very symmetric. However, about 100 RCs show
evidence of modest asymmetries, and about 25 RCs show severe global
asymmetries. In Table 1 we list, for each galaxy (ordered by alphanumeric
order; cols. [1] and [7]), the heliocentric systemic velocity (cols. [3] and
[9]) and the offset of the kinematic center from the photometric center (cols.
[2] and [8]), measured in arcsec along the major axis, along with the
inclination angle (cols. [5] and [11]).

It is important to have a reference scale for the optical size of each galaxy,
in order to assess the extension of each RC relative to that of the visible
matter. This is an essential step in order to evaluate the ability of RCs to
trace the
distribution of  matter out to where the DM becomes an important component. For
this purpose we have computed the radius, $R_{opt}$, encompassing 83\% of the
integrated light. (For an exponential disk this corresponds to 3.2
lengthscales, which in turn corresponds, for a Freeman disk, to the de
Vaucouleurs 25 $B$-mag/arcsec$^2$ photometric radius.) Specifically, from the
{\it I}--band CCD surface photometry of MFB, and correcting for inclination
(see Appendix), we have estimated $R_{opt}$ for each object (see cols. [4] and
[10] of Table 1).
\smallskip

\beginsection 3. The Rotation Curves.

We have identified 900 RCs which are symmetric, with reasonably low rms
internal scatter and reasonable high sampling density. These curves are shown
in Fig.1. (Filled/empty symbols represent approaching/receding sides; arrows
indicate optical radii.) This set of 900 RCs can be split into a subset of 80
{\it excellent} curves (Set A), and one of 820 {\it fair} curves (Set B).

The 80 RCs in Set A have the following properties, which ensure that they trace
well the gravitational potential in the region of interest and hence are well
suited for accurate and complete mass modelling: 1) the approaching and
receding arms are very symmetric; 2) the data are extended out to (at least)
$R_{opt}$; and 3) there are $\geq 30$ data points homogeneously distributed
with radius and (roughly) equally distributed between the two arms. For each
galaxy, from the folded RC we produce a smooth RC as follows. We smooth the
folded velocities by binning the $\leq N$ nearest data points contained within
a fixed maximum bin size $W$. The values used for $W$ are mostly
$0.050\,R_{opt}$ and $0.075\,R_{opt}$, and for $N$ are mostly 4 and 6. These
values are chosen for each RC according to its sampling density: however, the
profiles of the RCs do not change with other (reasonable) choices as can be
seen by comparing the smoothed RCs with the original folded ones. In each bin
we compute the average rotation velocity and its uncertainty. The resulting
smoothed curves are shown in Fig.2. Due both to the global smoothness of the
RCs and to the small uncertainty in each bin, the velocity fields are very well
traced by the binned data: as can be seen from Fig.2, a reasonable linearity in
the velocity profile extends for many curves through all radii, and for
virtually all curves between $0.4\,R_{opt}$ and $R_{opt}$ (in this latter
region only about 10\% of the data have been rejected owing to our adopted
requirement that $\rho > 0.35$ [see section 2]).

The 820 RCs in Set B fail, to various extents, at least one of the criteria
stated above and therefore may be not suitable for accurate and direct mass
modelling. However, they constitute a large database for those methods able to
recover the DM properties by needing less stringent requirement on the RC
quality. With this in mind, for each galaxy we estimate, from its folded RC,
the velocities at galactocentric radii corresponding to ${n \over 5} R_{opt}$
with $n=1, ..., 9$. Specifically, at each of these radii we compute the mean
velocity by averaging the data points over a bin of size $0.2\, R_{opt}$ and
centered on the reference radius. Then, to check stability, at the same radius
we define a pair of adjacent symmetric bins, each of size $0.15\, R_{opt}$; for
each bin of the pair we compute the average velocity; we then interpolate
between the pair of adjacent average velocities at the radius in question. Only
when the two values for the mean velocity obtained in this way are found to be
consistent with each other to within 10\%, do we consider our velocity estimate
reliable. In Table 2 we tabulate these estimated rotation velocities.
(Parenthesized velocities in Table 2 come from smooth eyeball
interpolation/extrapolation of the data points, for cases when we have deemed
such procedure safe.) Notice that, due to the very selection of this subset,
these velocities are often sparse and/or not extended out to $R_{opt}$. (For
convenience, in Table 2 we also report the corresponding velocities for RCs
belonging to Set A; and, for all galaxies, under the heading $V_{asympt}$ we
list the outermost estimated velocity [for radii $\geq 1.0\,R_{opt}$] as a
representative measure of the galaxy's asymptotic rotation velocity).

Finally, the 67 RCs of Set C have severe global asymmetries, large rms internal
scatter, insufficient sampling and/or large-scale deviations from circular
motion. These curves may be useful for studying non-axisymmetric disturbances
in spiral galaxies (e.g., bars, spiral arms). They are shown in Fig.3.

\beginsection 4. Conclusions.

We have presented the RCs of 967 spiral galaxies. By size, homogeneity, and
intrinsic quality of the individual curves, this sample constitutes by far the
best sample of RCs available to date. As such, it will offer a unique
opportunity for investigating in considerable depth the properties of dark
matter in galaxies, e.g. its radial distribution, total amount, scaling laws of
structural parameters. In a forthcoming paper (Persic, Salucci \& Stel 1995) we
will investigate the systematics of the 900 curves of Sets A and B and of their
underlying mass structure, and will attempt to clarify the implications of the
dark/visible mass interplay in spiral galaxies for current scenarios of galaxy
formation.
\smallskip

\noindent
{\it Acknowledgements.} We gratefully thank Don Mathewson for very helpful
exchanges and for his continuous encouragement throughout this work. We also
thank the referee, Vera Rubin, for a number of constructive comments on the
original manuscript.

%\vfill\eject
\vglue 1.truecm

\def\ref{\par\noindent\hangindent 20pt}
\centerline {\bf References }

\vglue 0.1truecm

\ref{Ashman, K.M. 1992, PASP, 104, 1109}
\ref{Cen, R., \& Ostriker, J.P. 1993, ApJ, 417, 415}
\ref{Evrard, A.E., Summers, F.J., \& Davis, M. 1994, ApJ, 422, 11}
\ref{Mannheim, P.D. 1993, ApJ, 419, 150}
\ref{Mathewson, D.S., Ford, V.L., \& Buchhorn, M. 1992, ApJS, 81, 413 (MFB)}
\ref{Navarro, J.F., \& White, S.D.M. 1994, MNRAS, 267, 401}
\ref{Persic, M., \& Salucci, P., \& Stel, F. 1995, in preparation}
\ref{Rubin, V.C., Burstein, D., Ford, Jr., W.K., \& Thonnard, N.
     1985, ApJ, 289, 81}
\ref{Stel, F. 1994, Laurea Degree thesis, University of Trieste}

\vfill\eject
\centerline{\bf APPENDIX}

In this Appendix we describe the estimation of the $I-$band visible radius for
the 967 galaxies of the present work. We define this radius as the radius
encompassing 83\% of the integral light, by analogy with the de Vaucouleurs
radius in the $B-$band (which statistically corresponds to 3.2 exponential
lengthscales and hence to the 83\% contour level of the total light).

By interpolation, from the individual surface brightness profiles (see MFB) we
have computed the radius, $R_{83}$, corresponding to 83\% of the total
$I-$light. To correct empirically for inclination effects, we have proceeded as
follows. First, we have taken into account a small inclination bias on the
deprojected velocities, present in the sample, of the form:
$$
{\rm log}\,\biggl<V_{83}({a \over b}) \biggr>^{\rm face-on}_{\rm sample}~ = ~
{\rm log}\,\biggl<V_{83}({a \over b}) \biggr>_{\rm sample}~ ~-~ (0.16 \pm 0.03)
	\times 	{\rm log}\,{a \over b}
\eqno(A1)
$$
where $V_{83} \equiv V(R_{83})$, and $a/b$ is the the major-to-minor axis ratio
(in the $I-$band, see column 4 in Table 1 of MFB) (for details see Stel 1994).
Then, we have assumed that
$$
{\rm log}\,R_{83}^{\rm true}~ = ~ {\rm log}\,R_{83}^{\rm obs} ~-~ C \times
	\biggl({\rm log}\,{a \over b}\biggr)^2\,,
\eqno(A2)
$$
where $C$ is a constant. (Of course we require null correction when $a/b=1$,
i.e. for face-on galaxies.) The constant $C$ has been determined by correlating
the residuals of the TF-like relationship log$\,R_{83}^{\rm obs}$ versus
log$\,V_{83})$ with the quantity $({\rm log}\,{a \over b})^2$. The whole
procedure has been iterated to convergence, yielding $C=0.16$ (see Fig.[A1]).
For cases with $a/b > 5$, we have adopted the correction corresponding to
$a/b=5$. In the main text $R_{83}^{\rm true}$ is referred to as the optical
radius $R_{opt}$.

\vfill\eject

\centerline{\bf Figure Captions.}

\bigskip

\noindent
{\bf Figure 1.} The 900 folded RCs of Sets A and B. Filled/empty symbols
denote approaching/receding sides. The triangles represent velocity points
which, though passing the S/N test, are discrepant and will not be used in
subsequent analyses. Velocities have been deprojected and corrected for the
$1/(1+z_{sys})$ relativistic redshift factor. For each galaxy, an arrow
indicates the optical radius, computed from individual $I$-band CCD surface
photometry (see Appendix for details).
\bigskip

\noindent
{\bf Figure 2.} The 80 smoothed RCs of Set A (see text for details). Vertical
arrows indicate optical radii.
\bigskip

\noindent
{\bf Figure 3.} The folded RCs of Set C (same symbols as for Fig.1). Only 61
curves are plotted. The RCs of the six galaxies 221-G2, 297-G27, 463-G21,
497-G32, 499-G18, and 545-G10 are not plotted because according to our
selection criterion on data quality those curves are void of significant data.

\bigskip

\noindent
{\bf Figure A1.} The correlation between the (binned) ratios of
apparent-to-corrected $R_{83}$ radii and the (log of the) apparent axial ratios
(all radii are in the $I$-band). The horizontal long-dashed line represents the
null correlation. The long/short-dashed line represents the slope of the
inclination correction we have used.
\bigskip

\bye